\documentclass[conference]{IEEEtran}
\IEEEoverridecommandlockouts
\usepackage{cite}
\usepackage{amsmath,amssymb,amsfonts}
\usepackage{algorithmic}
\usepackage{graphicx}
\usepackage{textcomp}
\usepackage{xcolor}
\usepackage{tabu}
\usepackage{array}
\usepackage{listings}
\usepackage{cuted}
\usepackage{float}
\usepackage[none]{hyphenat}
\usepackage{caption}
\def\BibTeX{{\rm B\kern-.05em{\sc i\kern-.025em b}\kern-.08em
    T\kern-.1667em\lower.7ex\hbox{E}\kern-.125emX}}

\begin{document}

\title{Enabling Secure and Effective Biomedical Data Sharing through Cyberinfrastructure Gateways
\thanks{Presented at Gateways 2020, Online, USA, October 12–23, 2020.
https://osf.io/meetings/gateways2020/}
}

\author{
    \IEEEauthorblockN{Shreya Goyal\IEEEauthorrefmark{1}, Saptarshi Purkayastha\IEEEauthorrefmark{1}, Tyler Phillips\IEEEauthorrefmark{2}, \\
    Robert Quick\IEEEauthorrefmark{3}, Alexis Britt\IEEEauthorrefmark{1}}
    \IEEEauthorblockA{
        \begin{tabular}{cc}
            \begin{tabular}{@{}c@{}}
                \IEEEauthorrefmark{1}Department of BioHealth Informatics, 
                \IEEEauthorrefmark{2}Department of Computer Science,
                \IEEEauthorrefmark{3}Pervasive Technology Institute\\
                \textsuperscript{‡} Indiana University,
                \textsuperscript{\**}\textsuperscript{†} Indiana University-Purdue University Indianapolis\\
                Indianapolis, Indiana 46202, USA. \\
                \{shregoya, saptpurk, phillity, rquick, almbritt\}@iupui.edu
            \end{tabular}
        \end{tabular}
    }
}

\maketitle

\begin{abstract}
Scientific cyberinfrastructures promise solutions to computational challenges with rich resources; they embrace collaborative workflows in which users can access and share scientific data and computing resources to perform research and education tasks, which catalyze scientific discovery. The Dynaswap project reports on developing a coherently integrated and trustworthy holistic secure workflow protection architecture for cyberinfrastructures which can be used on virtual machines deployed through cyberinfrastructure (CI) services such as JetStream. This service creates a user-friendly cloud environment designed to give researchers access to interactive computing and data analysis resources on demand. The Dynaswap cybersecurity architecture supports roles, role hierarchies, and data hierarchies, as well as dynamic changes of roles and hierarchical relations within the scientific infrastructure. Dynaswap combines existing cutting-edge security frameworks (including an Authentication Authorization-Accounting framework, Multi-Factor Authentication, Secure Digital Provenance, and Blockchain) with advanced security tools (e.g., Biometric-Capsule, Cryptography-based Hierarchical Access Control, and Dual-level Key Management). The security technologies have been developed and integrated with the Open Medical Record System for enhanced security purposes and delivering a secure scientific infrastructure, which allows researchers, educators, practitioners, and students to remotely access and share sensitive data, computing resources, and workflows with flexibility and convenience while also having the highest security and privacy protection. The CI is being validated in life-science research environments and in the education settings of Health Informatics.
\end{abstract}

\section{Introduction}
The Dynaswap project has developed an integrated and trustworthy holistic secure workflow protection architecture for cyberinfrastructures. The project is a CI gateway that allows different users to access and share proprietary data and computing resources to perform research and education tasks, and to bring about scientific discovery. Dynaswap is developed as an OpenStack Image on the JetStream Cloud Infrastructure \cite{stewart2015jetstream} supported by the National Science Foundation (NSF) for science and engineering research. This service creates a user-friendly cloud environment designed to give researchers access to interactive computing and data analysis resources on demand. The operational software environment that powers JetStream is OpenStack \cite{sefraoui2012openstack}. JetStream uses the OpenStack platform \cite{sefraoui2012openstack} to deploy the virtual machines as Infrastructure as a Service (IaaS) on the Internet2 and XSEDE networks. JetStream also uses the Globus Online system for secure file transfer between VMs such that SSH, X.509, and OpenID security protocols are available to all collaborating researchers of JetStream CI.

However, although widely used by biosciences researchers, the JetStream security features do not satisfy the strict requirements involving sensitive healthcare data and protected health information \cite{stewart2015big}. Gaps include a lack of preconfigured setup for privacy-preserving workflows, accessible deidentification, role-based differential access, flexible authentication, and authorization schemes. This results in substantial resistance from data owners and clinical collaborators to share data over JetStream VMs due to fear of data security violations \cite{farcas2013biomedical}. With a rise in precision health research, use of big data in healthcare, and demand for high-performance computing resources for data analytics and artificial intelligence, there is an urgent need for securing the CI through an easy-to-use, extensible, and robust cybersecurity architecture that can protect sensitive information being shared for scientific collaborations. The goal of the Dynaswap project is to incorporate existing cutting-edge security frameworks (including AAA, MFA, SDP, Blockchain) with advanced security tools (such as Biometric-Capsule, Cryptography-Based Hierarchical Access Control (CHAC), dual-level key management) to remotely access and share scientific data, computing resources, and workflows with flexibility and convenience while maintaining security and privacy.

The Dynaswap system architecture supports roles, role hierarchies, and data hierarchies, while ensuring protection with the latest security frameworks and tools. It also enables system administrators to dynamically change roles and hierarchical relationships within the cyberinfrastructure, allowing users to easily adjust data access in accordance with their needs. With these features, the gateway allows a wide range of users to collaborate by remotely sharing data, resources, and workflows in a flexible and secure environment. This architecture supports dynamic user (role) hierarchies and flexible associations between users and roles efficiently.

\section{Background}
The users of the source data have different roles, as illustrated in Figure 1 for a subset of the users. There are also different role levels for each type of user \cite{purkayastha2015dyadic}. For example, the nurses have multiple levels, including Certified Nursing Assistant, Registered Nurse, Licensed Practical Nurse, Nurse Practitioner, etc. In the same units, it is common for multiple nurses of different levels to work together for patient care. Cybersecurity challenges arise due to the complex access rights of these users, particularly when users may use medical records systems for clinical practice, research and training \cite{purkayastha2017implementation}. For instance, a physician should have unlimited access to all of the information of her/his own patients, but not full access to the information of other patients. A nurse should have only limited access rights to the information for those patients under her/his care according to the nurses' level. Clinical researchers can access information regarding their projects, while outside researchers can only access the information for their collaboration projects.

\begin{figure}[!ht]
  \centering
  \includegraphics[width=8cm, height=5cm]{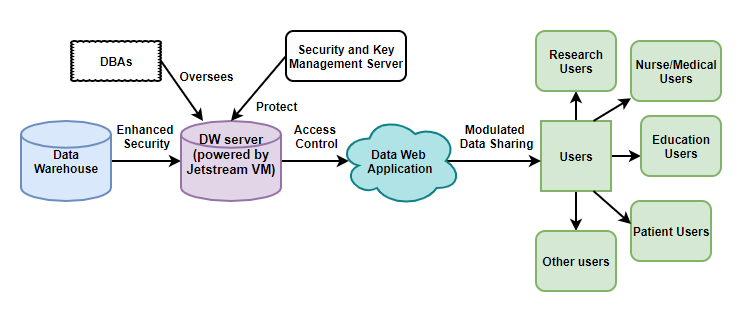}
  \caption{The research infrastructure on which the Dynaswap security architecture is applied and validated}
\end{figure}

The project categorize the user roles and allows fine-tuned data access using Cryptography-Based Hierarchical Access Control (CHAC) as shown in Figure 2.

\begin{figure}[!ht]
  \centering
  \includegraphics[width=8cm,height=5cm]{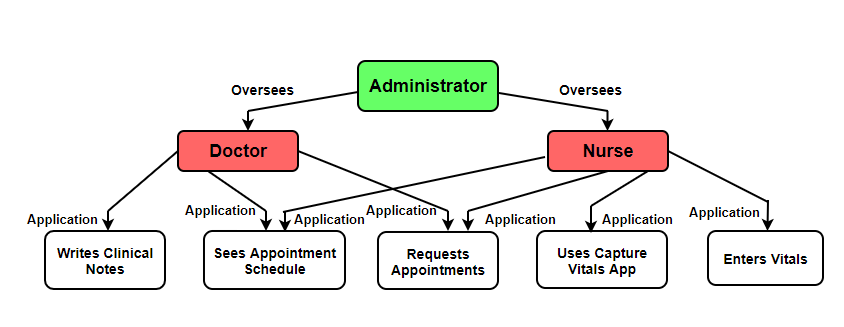}
  \caption{Cryptography-Based Hierarchical Access Control}
\end{figure}

It is assumed that the users can be grouped in the CI by roles and each group has multiple users. The role groups can be hierarchically related. Similarly, data can also be classified into different levels according to their sensitivities, privacy requirements, and/or structural composition relations and organized into hierarchies. Roles are associated with data, and users in a role group can access the corresponding data and all lower-level data in the hierarchy whose sensitivity is less than this data’s sensitivity. Moreover, due to the role hierarchy, a user was automatically granted all the sub-roles of the role in the role hierarchy. 

\section{Related Work}
CILogon \cite{basney2014cilogon} is a federated X.509 certification authority that has been in use for secure access to cyberinfrastructures such as the Extreme Science and Engineering Discovery Environment (XSEDE). CILogon supports multiple levels of assurance and customizable interfaces for user communities. The benefit of CILogon is that it relies on federated authentication for determining user identities when issuing certificates, hence enabling users to obtain certificates using their existing university or google identities. This eliminates the need/burden of the users to have to manage different identities and credentials for different services \cite{basney2014cilogon}. CILogon relies on the InCommon identity federation when issuing certificates \cite{basney2014cilogon}. 

The use of a Credential Store to manage the diverse resource credentials, map the end-user identities with community accounts, and enable the credential delegation without human interaction while creating a trusted multi-hop authorization is fairly common in many Science Gateways projects. The Credential Store is a secure database for storing authentication data with added utilities for performing delegation and key generation \cite{kanewala2014credential}. Delegation is defined as the mechanism of transferring end-user resource credentials to gateway middleware. If a delegation mechanism is available, the Credential Store uses it to transfer the credentials. Where there is no such mechanism, the Credential Store uses other approaches such as SSH key generation and manual persistence of credentials \cite{kanewala2014credential}.

The Science Gateway Platform (SciGaP) is an important CI gateway to advance scientific discovery that minimizes the net operating cost of Science Gateways\cite{marru2019experiences}. The SciGaP project's goal is to develop a robust platform with the ability to provide generic middleware functionalities and features required by the science gateways. The SciGaP infrastructure promotes sustainability through scaling. The paper provides various examples of Science Gateways being operated through multi-tenanted SciGaP Infrastructure for different field of sciences. Some of them includes SEAGrid, Ultrascan, PGA, InterACTWEL, Next GEN Thermo DB, Prostate Cancer Prediction \cite{marru2019experiences}. The Dynaswap project uses similar ideas for multi-tenanted infrastructure for deploying new VMs and still allowing users to share data, workflow and resources between each other.

Neuroscience Gateway (NSG) is an example of another science gateway infrastructure that makes neuroscience-specific computational tools conveniently available to the researchers and students \cite{carnevale2014neuroscience}. This portal allows the users to use the High Performance Computing (HPC) resources for modelling tasks with convenient web based interface that hides the technical details, enabling the users to focus on their research \cite{carnevale2014neuroscience}. In the Dynaswap project, we are able to provide a host of security tools to researchers, educators and clinical users of the OpenMRS medical records platform, such that those from low- and middle-income countries \cite{purkayastha2013post} are be able to access HPC resources through our gateway.

One of most popular gateways in the XSEDE community is the CIPRES (CyberInfrastructure for Phylogenetic RESearch) Science Gateway, designed to make the access easier and simpler to supercomputing resources \cite{miller2010creating}. This gateway allowed the researchers to get easy access to tools and resources for inferring evolutionary relationships. The availability of large and complex datasets have made it impossible to perform the analysis in local laptops. The CIPRES Science Gateway allows the to explore evolutionary relationships between species using supercomputers \cite{miller2010creating}. The iPlant Collaborative, another National Science Foundation (NSF) funded cyberinfrastructure (CI) project was launched in 2008 that allows the researchers to share their data, software tools, analysis pipelines, and best practices with their collaborators \cite{merchant2016iplant}. While originally developed for the plant science research community, iPlant has evolved to CyVerse and expanded to provide CI support across the life sciences \cite{merchant2016iplant}.

\section{Architecture}
An overview of the Dynaswap's architecture is shown in Figure 3. Many security, data processing, and data transfer features were employed to ensure data protection and support collaborative and dynamic scientific research and training.

\begin{figure}[!ht]
 \centering
  \includegraphics[width=8.5cm, height=5cm]{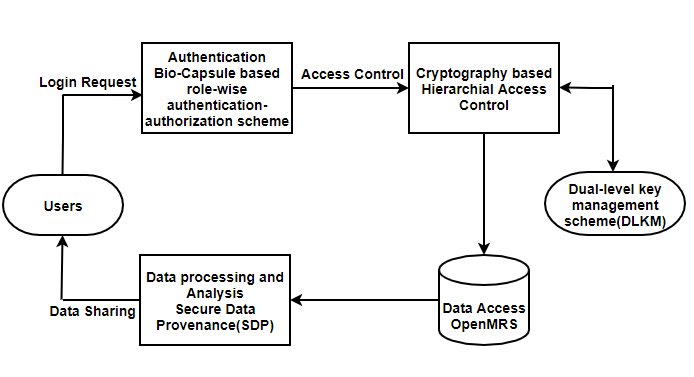}
  \caption{Dynaswap secure scientific infrastructure for sensitive data access and sharing}
\end{figure}

\subsection{Login Control}
The first step in a user attempting to gain access to the project architecture is for that user to login and be authenticated by the system. The method of confirming a user’s identity is a password-less Authentication-Authorization scheme (AuthN-AuthZ) based on a novel Bio-Capsule authentication model \cite{sui2013design}\cite{sui2011biometrics}. 

\subsubsection{Bio-Capsule Authentication Model}
Using the new Bio-Capsule (BC) authentication model \cite{sui2013design}\cite{sui2011biometrics}, BioCapsule-based role-wise authentication-authorization scheme was developed to realize role-based access control and data sharing \cite{phillips2017cancellable}. The BC generation process is through key extraction from a Reference Subject (RS - a random physical object or image) and the user’s biometric data. The key is then transformed and user and RS attributes were fused to generate the Bio-Capsule \cite{phillips2017cancellable}. This user-friendly method of authentication allows the user to use face identification to login and meet rigorous security standards \cite{jain2008biometric}.

\subsubsection{Authentication-Authorization Protocol}
Authentication verifies a user’s identity and authorization gives that user access to specific accounts. The Bio-capsule model lent some unique advantages to  Authentication-Authorization (AuthN-AuthZ) protocol. One advantage of the new BC model is that the RS (Reference Subject) can be shared by all users but each user receives a unique BC from his/her biometrics. Also, for two RSs, a user gets two totally different BCs using the same biometrics. This unique property enabled us to combine authentication and authorization together into one step.

\subsection{Access Control}
Once the user is authenticated by the system and authorized to login to a certain role, there were controls in place to determine what data the user can access based on the particular role that the user is authorized to perform in the system. Access to data within project architecture was defined by cryptography-based hierarchical access control (CHAC) \cite{atallah2009dynamic, chen2004efficient,ray2002cryptographic, tzeng2002time, zhong2002practical} and the system was enhanced by Dual-Level Key Management Scheme (DLKM), which enabled the system administrators to dynamically change roles in the hierarchical model to suit their needs.  

\subsubsection{Cryptography-Based Hierarchical Access Control (CHAC)}
Both roles and scientific data was organized in a hierarchical manner within the cryptography-based hierarchical access control scheme. If a user U with role R can access data D, U is automatically able to access all the data that are the children or descendants of D and in addition, to access all the data associated with the children or descendants of R. CHAC was used in this project to enable this kind of extended and fine-tuned access control. One important feature of our CHAC scheme is its ability to support dynamics, i.e., changes of a user’s role or changes to the hierarchy.

\subsubsection{Duel-Level Key Management Scheme (DLKM)}
An effective key management scheme was designed that supports dynamic changes of hierarchies and user groups at both the node level and user levels. Key management allows to effectively grant users access to certain data, according to the user’s role. The strength of this system includes its ability to dynamically change both the relationships between roles and users assigned to certain roles in the system. Also, the keys in the hierarchy have the property that any high-level key can derive its low-level keys but the reverse was not allowed.

\subsection{Data Access and Computation}
Once the user is authenticated and authorized to access a certain role, he or she must be able to view the appropriate data stored on the life sciences research server. Currently, Dynaswap works with the OpenMRS database, an open-source electronic medical records system \cite{purkayastha2019comparison}. However, the architecture could be applied to other databases where sensitive health information is stored. The project runs on an OpenMRS image (virtual machine) and allows users to securely access information stored in an OpenMRS database.

Once the data has been accessed researchers using this system would want to use a variety of computational methods to analyze and process the data. To protect the data, Secure Data Provenance (SDP) techniques were used, which aim to secure the data through its entire life-cycle \cite{muniswamy2006provenance,sultana2012secure,wang2012chaining}.

\section{Validation and Testing}
To make the secure workflow architecture a widely-used cyberinfrastructure, it needs to undergo rigorous testing and evaluation during its development and after completion. The project uses Common Vulnerability Scoring System (CVSS) v3 \cite{hanford2013common} as the vulnerability metrics for security assessment. The secured JetStream infrastructure is being assessed with rigorous testing by researchers and students in a semester-long graduate course in which a team of students simulate different user roles, are assigned various access privileges, and perform a variety of attacks against the cyberinfrastructure. Currently, the Usability testing for the BioCapsule authentication system is underway, and a survey has been sent to the students of the Privacy and Security course for the usability, security, and ease of use of the BioCapsule Authentication system implemented within the Dynaswap environment. Further, we are planning to include the testing and evaluation of the developed secure workflow and various clinical workflows. The evaluation design as a course assignment is shown in Tables 1 and 2 which includes the modules and CVSS metrics to be used for testing purposes. Table 2 provides an example of the evaluation of three types of workflows: login and access request workflow, data processing/analysis workflow, and data transfer among entities over internet workflow.
\begin{table}[!htbp]
\caption{Example course lab modules to test \& validate the secure workflows}
\fontsize{7}{8}\selectfont
\begin{tabular}{ 
|>{\raggedright\arraybackslash}p{3.0cm}|>{\raggedright\arraybackslash}p{5.1cm}| }
 \hline
 \textbf{Module lab name} & \textbf{Secure Workflow}  \\
 \hline
 Module 1: Patient records management & 1. Login to OpenMRS. Create patient cohort based on observations and diagnosis with a history of inflammatory
bowel disease and ulcerative colitis. \\ & 2. Create a report and then chart the distribution of patient age and gender. \\ & 3. Create an aggregation report based on the procedure code for the cohort. \\ 
 \hline
 Module 2: Principles of Security and Privacy: CAINA principles and Parkarian Hexad & 1. Login as a provider to get patient records for a patient who have not been seen by the provider (Confidentiality).\\& 2. Modify a patient record and try to change the data auditing tables to reflect change by another user (Integrity).\\& 3. Use an existing botnet, and target a server resource such that a provider from the network cannot access the patient record (availability).\\
 \hline
 Module 3: Data sharing and Protected Health Information & Create a cohort report based on patient symptom or condition and export C-CDA such that all HIPAA mandated 18 identifiers are removed from each patient record. \\
 \hline 
\end{tabular}
\end{table}

\begin{table}[!htbp]
\caption{Example of evaluation of the security-enhanced JetStream infrastructure}
\fontsize{7}{8}\selectfont
\begin{tabular}{ |>{\raggedright\arraybackslash}p{1.5cm}|>{\raggedright\arraybackslash}p{3.5cm}|>{\raggedright\arraybackslash}p{2.5cm}|  }
 \hline
 \textbf{Steps} & \textbf{Components tested} & \textbf{CVSS metric}  \\
 \hline
 Login & BC-based AuthN-AuthZ validation &  Attack Vector [N,A,L,P] \\
 \hline
 Access control & Role-wise AuthN-AuthZ and CHAC &  Privilege Required [N,L,H] \\
 \hline
 Access data & Role-wise AuthN-AuthZ, CHAC & Privilege Required [N,L,H] \\
 \hline
 Create/Compute &  CHAC & Privacy-preservation, Integrity [H,I,N] \\
 \hline
 Update &  SDP (Signature) &  Report Confidence [X,C,R,U] \\
 \hline
 Upload &  BC-based AuthN-AuthZ and CHAC &  Privilege Requires [N,L,H] \\
 \hline 
 Network transfer &  SDP (Enc., MAC) & Availability Req. [X,H,M,L]\\
 \hline 
\end{tabular}

 {\raggedright Notes: (1) Enc.: Encryption, MAC: Message Authentication Code.
(2) For metric values in [ H (high) and L (low)], each has its own meaning and empirical score.. E.g., for attack vector: N: Network (0.85), A:
Adjacent Network (0.62), L: Local (0.55) and P: Physical (0.2). The specification for all CVSS metric and possible values can be found in [27]. \par}
\end{table}

The CVSS score calculation is organized in two conceptually different groups, Exploitability metrics and Impact metrics. The specification for all CVSS metrics and possible values can be found in \cite{hanford2013common}. The results from the test will be analyzed and security measures will be improved as required. 

\section{Conclusion}
This paper discusses building a secure, holistic, fine-tuned, and multi-layered cybersecurity architecture that can be ported to multiple virtual machines on a popular CI like JetStream. This allows for scientific data/workflow so that researchers, educators, practitioners, and students can remotely access and share protected scientific data/workflow in a convenient yet finely-controlled manner. 

The novelty of our Dynaswap approach is the robust and secure workflow it offers using advanced security technologies. For securing the login process, a role-wise password-less Authentication-Authorization scheme (AuthN-AuthZ) based on a novel Biometric-Capsule is used. Also, the fine-tuned cryptography-based hierarchical access control scheme (CHAC) was developed and integrated with the OpenMRS. This ensures that only the authenticated legitimate users are able to login into the infrastructures and access the data.
For securing the data processing and analysis, use of SDP (secure digital provenance) is made to protect the scientific data and its provenance. The overall architecture makes sure that the stringent security measures are present at each step of the workflow. Another unique feature of the Dynaswap project is that it can accept and efficiently deal with user-role dynamics: the relationships among roles and among data can change and a user can join/be granted to, leave/be removed from, and change a role group according to changing application requirements and security needs.

The implemented secure architecture, when applied to healthcare and life-science infrastructure, will enable valuable clinical data to be broadly used by various users. Dynaswap will also help scale and protect NSF investments in national cyberinfrastructure such that they can be useful to healthcare projects. It also promotes the use of the rich, real clinical data in education. It would be interesting to see and evaluate, in the future, if the usability and performance of the Dynaswap architecture will be acceptable practically for the researchers in different context and countries. For example, forgetting/deleting ones own data as a patient or revoking the key if someone leaves the system ("right to erasure" in EU law) needs to be implemented in the future. 

\section{Acknowledgement}
The U.S. National Science Foundation supported this work under grant OAC-1839746. This work was made possible through a research allocation on the JetStream \cite{stewart2015jetstream} public cloud infrastructure and XSEDE resources.

\bibliographystyle{IEEEtran}
\bibliography{mybibfile}{}
\end{document}